\documentclass[3p]{elsarticle} 

\usepackage{hyperref}
\hypersetup{colorlinks=true, urlcolor=blue}

\usepackage{lineno,hyperref}
\modulolinenumbers[5]

\journal{Physics Letters A}

\usepackage{epsfig}
\usepackage{newlfont}
\usepackage{amssymb}
\usepackage{amsfonts}
\usepackage{amsmath}
\usepackage{bm}
\usepackage{subfigure}
\usepackage{times}
\usepackage{mathtools}
\usepackage{amsthm}
\usepackage[normalem]{ulem}

\newtheorem{theorem}{Theorem}

\theoremstyle{corollary}
\newtheorem{corollary}{Corollary}

\def\lsim{\mathrel{\rlap{\lower4pt\hbox{\hskip1pt$\sim$}}
    \raise1pt\hbox{$<$}}}                
\def\gsim{\mathrel{\rlap{\lower4pt\hbox{\hskip1pt$\sim$}}
    \raise1pt\hbox{$>$}}}                

\def \be{\begin{equation}}
\def \ee{ \end{equation} }

\begin{document}

\renewcommand*{\DefineNamedColor}[4]{%
   \textcolor[named]{#2}{\rule{7mm}{7mm}}\quad
  \texttt{#2}\strut\\}

\definecolor{red}{rgb}{1,0,0}
\definecolor{cyan}{cmyk}{1,0,0,0}

\title{Forbidden regimes in the distribution of bipartite quantum correlations due to multiparty entanglement}

\author[hri,hbni]{Asutosh Kumar}

\author[hri,vienna]{Himadri Shekhar Dhar}

\author[hri,iitp]{R. Prabhu}

\author[hri,hbni]{Aditi Sen(De)}

\author[hri,hbni]{Ujjwal Sen \corref{cor1}}
\ead{ujjwal@hri.res.in}

\cortext[cor1]{Corresponding author}

\address[hri]{Harish-Chandra Research Institute, Chhatnag Road, Jhunsi, Allahabad 211019, India}

\address[hbni]{Homi Bhabha National Institute, Anushaktinagar, Mumbai 400094, India}

\address[vienna]{Institute for Theoretical Physics, Vienna University of Technology,
Wiedner Hauptstraße 8-10/136, A-1040 Vienna, Austria} 

\address[iitp]{Department of Physics, Indian Institute of Technology Patna, Patna 800013, India}

\date{\today}

\begin{abstract}
%
Monogamy is a nonclassical property that limits the distribution of quantum correlation among subparts of a multiparty system. We show that monogamy scores for different quantum correlation measures are bounded above by functions of genuine multipartite entanglement for a large majority of pure multiqubit states. The bound is universal for all three-qubit pure states. We derive necessary conditions to characterize the states that violate the bound, which can also be observed by numerical simulation for a small set of states, generated Haar uniformly. The results indicate that genuine multipartite entanglement restricts the distribution of bipartite quantum correlations in a multiparty system.
\end{abstract}


\maketitle 


\section{INTRODUCTION}
\label{intro}

A fundamental property that distinguishes quantum correlation \cite{horodecki09,modidiscord}, which is a staple resource for exotic quantum information processing and computational tasks \cite{nielsen}, from classical 
correlation is its distribution among several parties in a quantum system. Classical correlation can be distributed among any number of  parties, with each pair attaining the maximum possible correlation.
However, for quantum correlations, there exist strong constraints on its sharability among the different parties of a multiparty system. Such constraints form the basis for the concept of monogamy of quantum 
correlations -- among three parties, if two possess maximal quantum correlations, they cannot have \emph{any} quantum correlation with the third party. All quantum correlation measures obey this qualitative monogamy relation while 
classical correlations do not \cite{monogamy}. 
In general, monogamy implies that if two systems are strongly correlated with respect to a nonclassical quantity, they can only be weakly correlated with respect 
to the same quantity to a third system. This property is satisfied by a host of nonclassical quantities including those related to Bell \cite{bell} and contextual inequalities \cite{context}, quantum steering witnesses \cite{steer, steer1},
and dense-coding capacities \cite{densehri}. The quantitative version of the monogamy constraint \cite{monogamy} is satisfied by entanglement monotones such as squared-concurrence, squared entanglement of formation,
squared-negativity, and squashed entanglement. 
The monogamy property is an important feature in quantum information theory \cite{nielsen} and in essence 
captures the ``trade-off'' between various quantifiers of quantum and classical properties \cite{koashi, christ}. It plays a significant role in the security of quantum key distribution \cite{qkd, qkd1} and in the considerations leading to quantum 
advantage in dense-coding \cite{horodense, horodense1}. Further, it has been used to characterize three-qubit genuinely multiparty entangled states \cite{viola, manab} and distinguish Bell-like orthonormal bases \cite{bell-like-asu}. 
For a recent review on monogamy of quantum correlations, see Ref.~\cite{mon-rev}.

From the perspective of quantum correlation, quantification of the monogamy property in multiparty systems is not very straightforward. This is in part due to the fact that quantum correlation shared among arbitrary parties in a multiparty 
system is not always computable, making the study of its distribution among several parties extremely difficult. However, there are ongoing efforts to overcome this constraint \cite{strong2}.
Further, characterization of both bipartite and multipartite quantum correlations in higher-dimensional mixed states is not well-developed \cite{horodecki09,modidiscord}. 
Nevertheless, various attempts have been made to systematically quantify the monogamy of quantum correlations. A seminal result was obtained in \cite{monogamy} for the monogamy of squared concurrence in three-qubit states. 
It was demonstrated that, for a three-qubit state, $\rho_{ABC}$, the sum of the squared concurrence between qubits $A$ and $B$, and that between qubits $A$ and $C$, is bounded above by the squared concurrence between qubit $A$ 
and the joint subsystem $BC$. Hence, the monogamy property was captured in the form of an inequality, known as the ``monogamy inequality''. An advantage of this inequality is that it is a multiparty property 
expressed in terms of bipartite quantum correlation measures which are well-understood at least for two-qubit systems.
The inequality was  shown to hold for squared concurrence in multiqubit quantum states 
\cite{monogamy2}.
However, it is not satisfied by some entanglement measures \cite{asuk,eofmono}, such as entanglement of formation and negativity. Further, information-theoretic measures of quantum correlation, such as quantum 
discord and quantum work-deficit, are also known to violate the monogamy inequality for three-qubits \cite{viola1, viola3}. Recent results on monogamy of quantum correlation have shown that 
the monogamy inequality is always satisfied for increasing powers of any quantum correlation measure \cite{newhri} or when large number of parties are considered \cite{newhri1}.

In this letter, we establish a relation between monogamy of quantum correlation, quantified by using the concept of ``monogamy score'' \cite{viola, manab, lightcone-prabhu}, and genuine multiparty entanglement measures for 
$n$-qubit pure states. The connection holds irrespective of the number of parties and is independent of the choice of the bipartite quantum correlation measure used in the conceptualization of monogamy.  
We show that for a large majority of pure multiqubit states, the monogamy score for a broad range of quantum correlation measures 
is upper bounded by a function of genuine multiparty entanglement in the system, as quantified by the generalized geometric measure (GGM) \cite{ggm, ggm2} (see also \cite{gm}).
We analytically show 
that the bound is universally satisfied for all pure three-qubit states \cite{lightcone-prabhu} and find conditions for its validity in arbitrary number of qubits
by identifying a set of necessary conditions to be satisfied in order to violate the bound for more than three qubits. 
Considering the squared concurrence and squared negativity as measures from the entanglement-separability paradigm, and quantum discord and quantum work-deficit as information-theoretic measures of quantum correlation, we numerically observe that these conditions are only satisfied for an extremely small set of $n$-qubit quantum states. In fact, by numerically generating 
random $4$- and $5$-qubit pure quantum states, using a uniform Haar distribution, we find that the bound is virtually never violated. 
The results show that the sharability of arbitrary bipartite quantum correlations in multisite quantum states, {irrespective of the number of parties, is nontrivially limited by the multiparty entanglement content of the states leading to forbidden regimes in the state space.}
{Another important aspect of the result is that the trade off between genuine multipartite entanglement and monogamy, establishes a potential upper bound on the monogamy score, which is an important quantity in the study of quantum correlations. However, the monogamy score is not easily computable for generic quantum states and measures of quantum correlation, and hence computable upper and lower bounds of the quantity provide the best option for evaluating the monogamy constraints on quantum correlations.}
We note that the monogamy inequality, as a physical characteristic, is dependent on the quantum correlation measure under consideration. A direct consequence of our work, as formulated more precisely in later sections, is that the monogamy score is now bounded above by a measure-independent quadratic or entropic function of GGM. Significantly, this allows us to readily obtain the upper bound of monogamy score, even for quantum correlation measures, such as distillable entanglement and entanglement cost, where the monogamy score is  either intractable or not computable. Along with a recent result giving a lower bound \cite{lowerbound} on the monogamy score, the present work provides an estimate of the constraints of monogamy in a quantum system.

This letter is organized as follows.
In Sec.~\ref{sec:definition}, we review the definitions of the monogamy score of a quantum correlation and the generalized geometric measure of an $n$-qubit state. In Sec.~\ref{bound}, we consider the multiparty entanglement bounds on monogamy score in terms of the genuine multiparty entanglement. In Sec.~\ref{anal}, we analytically show how the bound is satisfied for a host of $n$-qubit ($n > 3$) symmetric and many-body ground states. In Sec.~\ref{num}, we numerically find that the bound is satisfied for randomly generated four- and five-qubit
quantum states. We conclude in Sec.~\ref{conclude}.  

\section{Definitions}
\label{sec:definition}

In this section, we discuss definitions of monogamy of quantum correlations and genuine multipartite entanglement. For an $n$-qubit pure state, $|\Psi\rangle$, 
the monogamy inequality \cite{monogamy} of a bipartite quantum correlation measure, $\cal{Q}$, with respect to a nodal qubit (say $j$), can be written as
\begin{eqnarray}
\cal{Q}(\rho_{j:rest}) &\geq & \sum_{k\neq j} \cal{Q}(\rho^{(2)}_{jk}),
\label{ineq}
\end{eqnarray}
where $\cal{Q}(\rho_{j:rest})$ is the bipartite quantum correlation between the nodal qubit and the rest of the qubits taken as a single party, and $\cal{Q}(\rho^{(2)}_{jk})$ is the bipartite 
quantum correlation measure between the nodal qubit $j$ and the qubit $k$, obtained from  the two-qubit reduced density matrix $\rho^{(2)}_{jk}$. A multipartite state which satisfies the monogamy relation is said to be monogamous, and otherwise it is non-monogamous. Quantum correlation measures that satisfy the above inequality for all multipartite states are termed as monogamy-satisfying or monogamous. Entanglement monotones such as squared concurrence, squared entanglement of formation, and squared negativity are known to be monogamous, whereas entanglement of formation and logarithmic negativity, along with information-theoretic quantum correlation measures such as quantum discord and quantum work-deficit are, in general, non-monogamous. The definitions of some of these measures are provided in the Appendix.

Rewriting Eq.~(\ref{ineq}), one can define monogamy score as
\begin{equation}
\delta_{j}^\cal{Q}(|\Psi\rangle) = \cal{Q}(\rho_{j:rest}) - \sum_{k\neq j} \cal{Q}(\rho^{(2)}_{jk}),
\label{score}
\end{equation}
with qubit $j$ as the node. It is non-negative for all the states that satisfy Eq.~(\ref{ineq}), while for non-monogamous states, it possess negative values. On the other hand,  all monogamous quantum correlation measures have a positive monogamy score, for all multipartite states. An important aspect in defining both the monogamy inequality and the score is the role of the nodal qubit. For $n$ = 3, the monogamy score with respect to squared concurrence is independent of the choice of a nodal qubit \cite{monogamy}. However, for other quantum correlation measures and for $n >$ 3, this invariance is lost for general multiqubit states. 
A logical step is to consider the monogamy score after finding the minimal value across all choices of nodal qubits, i.e., $\delta^\cal{Q}(|\Psi\rangle) = \min_{j \in [1,n]}[\left\{\delta_{j}^\cal{Q}(|\Psi\rangle)\right\}]$, 
where $j \in [1,n]$ indicates that $j$ is chosen from among $\{1,2,\ldots,n\}$. 

In recent years, there have been several attempts to characterize multipartite entanglement in terms of geometric measures \cite{ggm, ggm2, gm}, relative entropy \cite{relative}, characteristic polynomials of reduced states \cite{poly1}, and probabilistic distribution of bipartite entanglement \cite{Facchi}. 
In the present work, we consider the 
genuine multiparty entanglement of an $n$-qubit pure state, $|\Psi\rangle$, which can be conceptualized by using the \emph{generalized geometric measure} (GGM) \cite{ggm, ggm2} (cf. \cite{gm}). $|\Psi\rangle$ is said to be
genuinely multipartite entangled if it cannot be expressed as a product across any bipartition of the state. The Greenberger-Horne-Zeilinger (GHZ) and the W states are the quintessential 
examples of genuinely multipartite entangled states. The GGM ($\cal{G}$) of the state $|\Psi\rangle$ can be reduced to
\begin{equation}
\cal{G}(|\Psi\rangle) = 1 - \max_{\left\{|\Phi\rangle\right\}}|\langle \Phi|\Psi \rangle|^2, 
\end{equation}
where the maximization is performed over the set of states $\left\{|\Phi\rangle\right\}$ that are not genuinely multiparty entangled. From the definition, it follows that the quantity $\cal{G}(|\Psi\rangle)$ vanishes for all 
states that are biseparable across any partition and is non-zero otherwise. Further, it is a valid entanglement monotone that is non-increasing under local operations and classical communication. The optimization in defining 
GGM can be simplified in terms of the maximization of the Schmidt coefficients across all possible bipartitions, allowing the quantity to be calculated for arbitrary pure states in arbitrary dimensions. In terms of the Schmidt 
coefficients, the GGM for $|\Psi\rangle$ can be defined as  \cite{ggm, ggm2}
\begin{equation}
\cal{G}(|\Psi\rangle) = 1 - \max\left\{\lambda^2_{A:B}| A \cup B = \left\{1,2,...,n\right\}, A \cap B = \emptyset\right\},
\label{GGM}
\end{equation}
where, $\lambda_{A:B}$ is the maximal Schmidt coefficient across the bipartition $A:B$ of the state $|\Psi\rangle$. This allows one to compute $\cal{G}(|\Psi\rangle)$ in terms of the eigenvalues of its different reduced density 
matrices.

\section{Monogamy score and genuine multiparty entanglement}
\label{bound}

In this section, we connect the monogamy score with genuine multiparty entanglement measure. In particular, we show that the monogamy score of a quantum correlation measure for any multiqubit pure state is upper bounded by the genuine multiparty entanglement of the state, quantified using the generalized geometric measure \cite{ggm, ggm2}. 
{The significance of GGM as a potential measure is due to its analytical and computational accessibility. GGM is a computable measure of genuine multipartite entanglement in pure quantum states of arbitrary number of parties in arbitrary dimensions. Meanwhile, monogamy score is a measure of ``residual'' quantum correlations \cite{monogamy, manab}, not captured by bipartite correlations, and has traditionally been used as an intuitively satisfactory measure of multipartite quantum correlations. The trade-off between the two multiparty quantities is physically interesting as it leads to definite regimes in the state-space that are inaccessible or forbidden.}

Let us consider an $n$-qubit pure state $|\Psi\rangle$. The corresponding $k$-qubit reduced states are given by $\rho^{(k)}$ = $\mathrm{Tr}_{n-k}(|\Psi\rangle\langle\Psi|)$, where $n-k$ parties 
have been traced out. From the definition of GGM, we know that $\cal{G} = 1 - \max_{ k \in [1,n/2]}  \left[\left\{\xi_{m}(\rho^{(k)})\right\}\right]$, where $\left\{\xi_{m}(\rho^{(k)})\right\}$ is the set of maximum eigenvalues corresponding to 
all possible $k$-qubit reduced states of the $k:n-k$ bipartitions, for $k$ ranging from $1$ to $n/2$.
Let us now establish the connection between GGM and monogamy of bipartite measures ${\cal Q}$.

\begin{theorem}
{For all multiqubit pure states, $|\Psi\rangle$, the monogamy score, $\delta^\cal{Q}(|\Psi\rangle)$, of a quantum correlation measure, $\cal{Q}$, is bounded above by a 
function of the generalized geometric measure, $\cal{G}(|\Psi\rangle)$, provided the maximum eigenvalue in obtaining $\cal{G}$ emerges from a single-qubit reduced density matrix.}
\label{p1}
\end{theorem}
\noindent\textbf{Proof}: {
Let $a =\max \{\xi_{m}(\rho^{(1)})\}$ be the maximum eigenvalue
corresponding to all possible single-qubit reduced density matrices, $\rho^{(1)}$ of
$|\Psi\rangle$, 
obtained from, say qubit $j$. The monogamy score for node $j$ is given by $\delta_{j}^\cal{Q} = \cal{Q}(\rho_{j:\mathrm{rest}}) - \sum_{k\neq j} \cal{Q}(\rho^{(2)}_{jk})$. Therefore, one obtains $\delta_{j}^\cal{Q} \leq \cal{Q}(\rho_{j:\mathrm{rest}})$.
Since $\cal{Q}$ is local unitary invariant, the quantity $\cal{Q}(\rho_{j:\mathrm{rest}})$ is a function of the maximum eigenvalue $a$, say $f^{\cal{Q}}(a)$. Since, the maximum eigenvalue in obtaining the generalized geometric measure $\cal{G}$ emerges from a single-qubit reduced density matrix, we have $\cal{G}$ = $1 - a$. Thus we have $f^{\cal{Q}}(a)$ = $\cal{F}^{\cal{Q}}(\cal{G})$, which gives us the bound, 
$
\delta_{j}^\cal{Q} \leq f^{\cal{Q}}(a) = \cal{F}^{\cal{Q}}(\cal{G})$.
Now the monogamy score, $\delta^\cal{Q}$, is defined as the 
minimum score over all possible nodes. Hence, $\delta^\cal{Q} \leq \delta_{j}^\cal{Q}$, and thus we obtain an upper bound on the monogamy score in terms of a function of generalized geometric measure, as given 
by $\delta^\cal{Q}(|\Psi\rangle) \leq  \cal{F}^{\cal{Q}}(\cal{G}(|\Psi\rangle))$.}
\hfill\(\blacksquare\)
{From the proof of Theorem~\ref{p1}, it may appear that the $n-1$ bipartite correlation terms, $\cal{Q}(\rho^{(2)}_{jk})$, are not given due privilege. However, this is not the case. This is observed, for example, in Figs.~\ref{figure1}-\ref{figure3}, where the region just above the curved boundaries contain representatives of a large number of multiparty states for which $\sum_{k \neq j} \cal{Q}(\rho^{(2)}_{jk}) \neq$ 0. Only on the boundary, the bipartite terms are vanishing. As we move away from the boundary, the bipartite contributions come into the picture. The bound in Theorem 1, states that only the region on and above the boundary is populated.}

Let us consider the form of the function $\cal{F}^{\cal{Q}}(\cal{G}(|\Psi\rangle))$ in Theorem~\ref{p1} for a set of quantum correlation measures that reduce to the von Neumann entropy for pure states. Such measures include entanglement of formation \cite{eof}, squashed entanglement \cite{koashi, christ}, distillable entanglement \cite{distil-ent}, entanglement cost \cite{cost},  relative entropy of entanglement \cite{relative}, quantum discord \cite{discord1, discord2}, and quantum wok-deficit \cite{qwd1}. For these measures, the quantity $\cal{Q}(\rho_{j:\mathrm{rest}})$ is equal to $S(\rho_j^{(1)})$. If $a$ is the maximum eigenvalue of $\rho_j^{(1)}$, then 
$S(\rho_j^{(1)})$ = $-a\log_2 a-(1-a)\log_2(1-a)$ = $h(\cal{G})$, where $\cal{G}$ = $1-a$. Therefore, $\cal{F}^{\cal{Q}}(\cal{G}(|\Psi\rangle))$ = $h(\cal{G})$. For all these measures, a scatter diagram similar to Figs.~\ref{figure1}-\ref{figure3}, plotted in this instance for quantum discord ($\cal{D}$) and quantum work-deficit ($\Delta$), shall show that the monogamy score is bounded by the well-defined curve of the von Neumann entropy, i.e., $\delta^\cal{Q}(|\Psi\rangle) \leq  S(\rho_j^{(1)})$ = $h(\cal{G})$. 
An important implication of Theorem~\ref{p1} is that the above bound is valid for even those quantum correlation measures that can not be explicitly computed for arbitrary states using any analytical or numerical methods, such as distillable entanglement, entanglement cost, and relative entropy of entanglement. The theorem implies that any possible value of these measures will always result in monogamy scores that lie on or above the boundary. Moreover, GGM is a well-defined measure of genuine multipartite entanglement, which is monotonically non-increasing under local operations and classical communication, and thus provides a strong physical connection between the monogamy score bound and multiparty entanglement.

{Moreover, one can also consider measures of quantum correlation that reduce to functions of the determinant of local density matrices, such as negativity, logarithmic negativity, and concurrence.  
Let us consider the squares of concurrence ($\cal{C}^2$) and negativity ($\cal{N}^2$)
as such quantum correlation measures. 
For $\cal{C}^2$ and $\cal{N}^2$, the quantity $\cal{Q}(\rho_{j:\mathrm{rest}})$ is equal to $4 \det(\rho^{(1)}_j)$ and $\det(\rho^{(1)}_j)$, respectively. 
Therefore, we obtain $\cal{Q}(\rho_{j:\mathrm{rest}})$ = $za(1-a)$
= $zg(\cal{G})$, where $z$ = 4 and 1 for $\cal{C}^2$ and $\cal{N}^2$, respectively. For all such measures, the bound on monogamy score in well-defined by the function $zg(\cal{G})$.}
We note that the form of the function, $\cal{F}^{\cal{Q}}(\cal{G}(|\Psi\rangle))$, is dependent on the type of the quantum correlation measure, which is reminiscent of the fact that the monogamy inequality is a measure-specific characteristic. Interestingly, as shown for distillable entanglement and entanglement cost, this does not hinder the computation of the bound for cases where the monogamy score, in general, can not be computed. 

%
The applicability of the bound obtained in Theorem~\ref{p1} is limited, in the sense that it is only valid for genuinely multipartite entangled pure states for which
the maximum Schmidt coefficient contributing to the GGM of the state comes from the $j$:$rest$ bipartition, where $j$ is the single qubit. However, as we shall observe in the following analysis, the 
bound in Theorem \ref{p1} holds for a large number of randomly generated multiqubit states. Numerical studies, for a few qubits, show that there is only a small fraction of randomly generated states for which the  maximal Schmidt coefficient comes from  bipartitions other than those containing a single qubit.
Nevertheless, the results of Theorem \ref{p1} can be extended to other states with specific restrictions.\\

\noindent{\bf Proposition 1.} {\em For $n$-qubit pure states $|\Psi\rangle$, where $n >$ 3 and the maximum eigenvalue in calculating the generalized geometric measure $\cal{G}(|\Psi\rangle)$ emerges from a reduced 
density matrix, $\rho^{(k)}$, with $k \neq 1$, the upper bound of monogamy score, $\delta^\cal{Q}(|\Psi\rangle)$, of a quantum correlation, $\cal{Q}$, is a function $\cal{F}^{\cal{Q}}(\cal{G})$ of $\cal{G}$, provided the function $\cal{H}^{\cal{Q}}(|\Psi\rangle) \ge$ 0.}\\

\noindent\textbf{Proof}: {From Theorem~\ref{p1}, we know the monogamy score satisfies the relation, $\delta^\cal{Q} \le f^{\cal{Q}}(a)$, where $a$ = $\{\xi_{m}(\rho^{(1)})\}$ is the maximum eigenvalue corresponding to all possible single-qubit reduced density matrices. However, in this case, the generalized geometric measure, $\cal{G}$ = $1 - b$, where $b$ = $\{\xi_{m}(\rho^{(k)})\}$, $\forall~ k \neq 1$, is the maximum eigenvalue corresponding to all possible non-single qubit bipartitions. Hence, the premise implies that $b > a$ and we have $f^{\cal{Q}}(a) \neq \cal{F}^{\cal{Q}}(\cal{G})$.
Therefore, though we have $\delta^\cal{Q} \leq  f^{\cal{Q}}(a)$ it does not necessarily imply 
$\delta^\cal{Q} \leq  \cal{F}^{\cal{Q}}(\cal{G})$. 
Let us define the quantity, $\beta$ = $b - a > 0$. Now, $f^{\cal{Q}}(a)$ = $f^{\cal{Q}}(b-\beta)$. Expanding the above expression, one can show that $f^{\cal{Q}}(b-\beta)$ = $f^{\cal{Q}}(b) - \cal{R}^{\cal{Q}}(b,\beta)$, with $\cal{R}^{\cal{Q}}(b,\beta)$ being function of $b,\, \beta$ and ${\cal Q}$ which can be different for various quantum correlation measures, as shown in Table~{\ref{table1}}. Since, $\cal{G}$ = $1-b$, the bound on 
monogamy score, using $f^{\cal{Q}}(b)$ = $\cal{F}^{\cal{Q}}(\cal{G})$, can be written as
\begin{eqnarray}
\delta^\cal{Q} &\leq&  f^{\cal{Q}}(b) - \cal{R}^{\cal{Q}}(b,\beta)
=  \cal{F}^{\cal{Q}}(\cal{G}) - \cal{R}^{\cal{Q}}(b,\beta).
\label{scored}
\end{eqnarray}
Therefore, we obtain the bound $\delta^\cal{Q} \leq  \cal{F}^{\cal{Q}}(\cal{G})$, provided $\cal{R}^{\cal{Q}}(b,\beta) \geq$ 0. However, $a \ge 1/2$ and for $b > a \ge 1/2$, 
it can be easily shown that the function $\cal{R}^{\cal{Q}}(b,\beta)$ is always negative. For example, quantum correlation measures where $f^\cal{Q}(a)$ = $zg(a)$ or $h(a)$, the expressions $1-2b+\beta$ and $h(b)-h(b-\beta)$, respectively (see Table~\ref{table1}), are negative for $\beta > 0$ and $1 \ge b \ge b-\beta$.
Hence, to satisfy the bound, {we need to consider the contribution of additional terms in the monogamy inequality. In this instance, we note that the states considered in the Proposition are reduced compared to those in Theorem~\ref{p1}. Let us consider, the relation for the monogamy score in Eqs.~\ref{score} and \ref{scored}, which can be rewritten as,
\begin{eqnarray}
\delta^\cal{Q} &=&  f^\cal{Q}(a) - \sum_{k\neq l} \cal{Q}(\rho^{(2)}_{lk}) =
f^{\cal{Q}}(b) - \cal{R}^{\cal{Q}}(b,\beta) - \sum_{k\neq l} \cal{Q}(\rho^{(2)}_{lk})\\
&=&  \cal{F}^{\cal{Q}}(\cal{G}) - \cal{H}^{\cal{Q}}(|\Psi\rangle), 
\label{bound-p1}
\end{eqnarray}
where from Theorem~\ref{p1}, $f^\cal{Q}(a)$ = $\cal{Q}(\rho_{j:\mathrm{rest}})$, and the function $\cal{H}^{\cal{Q}}(|\Psi\rangle)$ is given by
\begin{equation}
\cal{H}^{\cal{Q}}(|\Psi\rangle) = \sum_{k\neq l} \cal{Q}(\rho^{(2)}_{lk}) + \cal{R}^{\cal{Q}}(b,\beta),
\end{equation}
where $l$ corresponds to the node for which the monogamy score is minimal. We note that the quantities $b$, $\beta$, and $\cal{R}^{\cal{Q}}$ are independent of the node $l$.
Therefore, we once again obtain the bound, $\delta^\cal{Q} \leq  \cal{F}^{\cal{Q}}(\cal{G})$, provided $\cal{H}^{\cal{Q}}(|\Psi\rangle) \geq 0$\footnote{The condition $\cal{H}^{\cal{Q}}(|\Psi\rangle) \geq 0$, implies that the distribution of bipartite entanglement between the nodal party, $l$, and each of the parties, $k$, given by $\sum_{k\neq l} \cal{Q}(\rho^{(2)}_{lk})$, must be larger than the absolute value of $\cal{R}^{\cal{Q}}(b,\beta)$ for the bound to be satisfied.}. }}
\hfill\(\blacksquare\)\\

At this point, we note that for a system composed of a large number of qubits or arbitrary quantum correlation measures, the function $\cal{H}^{\cal{Q}}(|\Psi\rangle)$, is in general, not easily accessible as it requires explicit calculation of the terms $\cal{Q}(\rho^{(2)}_{lk})$. However, for specific states and measures, including symmetric states, or for small number of qubits these quantities can be efficiently computed. In Section~\ref{anal}, we analytically estimate the quantities in Proposition~1, for several important classes of states, viz., the $n$-qubit Dicke states, the superposition of generalized GHZ and W states, and ground states of interesting physical systems, such as quantum spin-$1/2$ lattices. Also, in Section~\ref{num}, we numerically estimate these quantities for randomly generated Haar-uniform four and five qubit pure states.   
All these states provide important instances where the upper bound on monogamy score in terms of the GGM are exemplified. Proposition~1 thus provides the generic mathematical apparatus to estimate the bound on monogamy score. Analytical and numerical analyses of symmetric and random pure states show that the fraction of states that satisfy the above conditions, and thus may violate the bound on monogamy score, is extremely small. Table~\ref{table2} indicates the percentages of states that satisfy each of these conditions for different classes of states. It is evident that a large majority of four and five qubit states satisfy the bound on monogamy score.

\begin{table}[t]
\begin{center}
\begin{tabular}{|c|c|c|}
\hline
\textbf{${\cal Q}$}     & \textbf{$f^{\cal{Q}}(b)$}  & \textbf{$\cal{R}^{\cal{Q}}(b,\beta)$} \\ \hline 
${\cal C}^2$           & $4b(1-b)$ & $4\beta(1-2b+\beta)$ \\ \hline
${\cal N}^2$           & $b(1-b)$  & $\beta(1-2b+\beta)$  \\ \hline
${\cal D}$, $\Delta$  & $h(b)$    & $h(b)-h(b-\beta)$ \\ \hline
\end{tabular}
\caption{Expressions of \(\cal{F}^{\cal{Q}}(b)\) and \(\cal{R}^{\cal{Q}}(b,\beta)\) for concurrence-square (${\cal C}^2$), negativity-square (${\cal N}^2$), quantum discord (${\cal D}$) and quantum work-deficit ($\Delta$), where 
$h(x)=-x\log_2 x-(1-x)\log_2(1-x)$, is the binary Shannon entropy.}
\label{table1}
\end{center}
\end{table}
%

\begin{table}[htb]
\begin{center}
\begin{tabular}{|c|c|c|c|c|c|}
\hline
State & $\beta$  & $\cal{H}^{\cal{C}^2}$ & $\cal{H}^{\cal{N}^2}$ & $\cal{H}^{\cal{D}}$ & $\cal{H}^{\Delta}$ \\ \hline 
~~~~$|G^1\rangle$~~~~                & ~~ 99.87~~       & ~~0.007~~       & ~~~~0~~~~         & ~~~~0~~~~       & ~~~~0~~~~                       \\ \hline
$|G^2\rangle$                & 91.65           & 0                     & 0                     & 0                   & 0                       \\ \hline
$|G^3\rangle$                & 57.04            & 0                     & 0                     & 0                   & 0                       \\ \hline
$|G^4\rangle$                & 97.58           & 0                     & 0                     & 0                   & 0                       \\ \hline
$|G^5\rangle$                & 60.77            & 0                     & 0                     & 0                   & 0                       \\ \hline
$|G^6\rangle$                & 4.97            & 0                     & 0                     & 0                   & 0                       \\ \hline
$|\psi^4_{sym}\rangle$       & 6.37            & 0                     & 0                     & 0                   & 0                       \\ \hline
$|\psi^5_{sym}\rangle$       & 0.303           & 0                     & 0                     & 0                   & 0                       \\ \hline
$|\psi^4_{gen}\rangle$       & 4.44            & 0                     & 0                     & 0                   & 0                       \\ \hline
$|\psi^5_{gen}\rangle$       & 0.26            & 0.12                  & 0.125                 & 0                   & 0                        \\ \hline
\end{tabular}
\caption{Percentages of states that satisfy the necessary conditions, $\beta > 0$ and $\cal{H}^{\cal{Q}}(|\Psi\rangle) <$ 0. We randomly generate $2.5 \times 10^5$ states belonging to different classes using uniform Haar distributions. States that simultaneously satisfy all three inequalities may violate the monogamy bound. $|G^i\rangle$ ($i$ = 1 to 6) are parameterized  inequivalent classes under stochastic local operation and classical communication (SLOCC) for four-qubit states defined in \cite{verstraete-9classes,chterental}. $|\psi^k_{sym}\rangle$ and 
$|\psi^k_{gen}\rangle$ are randomly generated symmetric and general $k$-qubit states, respectively. See Appendix for the definitions of the states.}
\label{table2}
\end{center}
\end{table}
%
For symmetric $n$-qubit pure states, $|\Psi\rangle$, as encountered repeatedly in the following sections, the expression for $\cal{H}^{\cal{Q}}(|\Psi\rangle)$ simplifies significantly, as all the $n-1$ bipartite quantum correlation contributions, $\cal{Q}(\rho^{(2)}_{jk})$, are equivalent. Hence, the expression for  $\cal{H}^{\cal{Q}}(|\Psi\rangle)$ reduces to
\begin{equation}
\cal{H}^{\cal{Q}}(|\Psi\rangle) = \cal{R}^{\cal{Q}}(b,\beta) + (n-1)\cal{Q}(\rho^{(2)}_{12}).
\label{bound2-p1}
\end{equation}
Hence, to violate the bound on monogamy score, an $n$-qubit pure state $|\Psi\rangle$, where $n >$ 3, must simultaneously satisfy the following necessary conditions: $\beta > 0$ and $\cal{H}^{\cal{Q}}(|\Psi\rangle) < 0$. 

An important question to consider is -- \textit{When does a multiparty state violate the bound on monogamy score?} From previous discussions, it is clear that $\cal{H}^{\cal{Q}}(|\Psi\rangle) < 0$ ensures that the state will not satisfy Proposition~1. Moreover, it has also been established that $\cal{H}^{\cal{Q}}(|\Psi\rangle)$ may not always be analytically accessible, or even numerically tractable, for all states and quantum correlation measures. Hence, it is impossible to analytically characterize a multiparty state that may violate the bound in Proposition~1. 
However, one can construct examples of genuinely multiparty separable states, to
highlight certain properties of states that may violate the bound on monogamy score.
For instance, consider the six-qubit state, $|\Psi\rangle$ = $|\psi_g\rangle \otimes |\psi_g\rangle$, where $|\psi_g\rangle$ = $\frac{1}{2}(|000\rangle + |111\rangle)$ is the 3-qubit Greenberger-Horne-Zeilinger (GHZ) state. First, note that the state is not genuinely multiparty entangled and hence, $\mathcal{G}(\Psi)$ = 0 ($b$ = 1). Interestingly, for a GHZ state, $\cal{Q}(\rho^{(2)}_{lk})$ = 0, $\forall~ l,k$. It implies that choosing any qubit as the nodal site, we obtain, $\cal{H}^{\cal{Q}}(|\Psi\rangle) < 0$, since $\cal{R}^{\cal{Q}}$ = $-1$. Therefore, the state $|\Psi\rangle$ can violate Proposition~1. This is true since the monogamy score, for the measure $\mathcal{C}^2$, is equal to 1, for $|\Psi\rangle$. Thus, $\delta^\mathcal{Q} > \mathcal{G}$. We note that the crux of the question lies in the quantity, $\cal{Q}(\rho^{(2)}_{lk})$ and its relation to the function, $-\cal{R}^{\cal{Q}}$ = $f^\mathcal{Q}(b)-f^\mathcal{Q}(a)$. For states with low $\mathcal{G}$ (high $b$) and low aggregate of $\cal{Q}(\rho^{(2)}_{lk})$ , the term $-\cal{R}^{\cal{Q}}$ can dominate in $\cal{H}^{\cal{Q}}(|\Psi\rangle)$, leading to the violation of Proposition~1. 

\begin{corollary}
For all three-qubit pure states $|\Psi\rangle$, the monogamy score, $\delta^\cal{Q}(|\Psi\rangle)$, is upper bounded by a function of the generalized geometric measure, $\cal{G}(|\Psi\rangle)$, for all quantum correlation measures, $\cal{Q}$.
\label{col2}
\end{corollary}

\noindent\textbf{Proof}: For any three-qubit pure state $|\Psi\rangle$, for all bipartitions, the relevant reduced density matrices are the 
single-qubit reduced density matrices $\{\rho^{(1)}\}$. Hence, the maximum eigenvalue contributing to the generalized geometric measure, $\cal{G}(|\Psi\rangle)$, always comes from $\{\rho^{(1)}\}$, thus satisfying the premise of Theorem~\ref{p1}.
\hfill\(\blacksquare\)

{We note that the result in Corollary~\ref{col2}, which is a limited case of the generalized $n$-qubit statement presented in Theorem~\ref{p1} of our study, was formally shown in \cite{lightcone-prabhu} for three-qubit states. Since, the results presented therein are restricted to three-party systems, it does not contain any information about those pure states where the maximum eigenvalue contributing to the calculation of GGM does not arise from the $1$:$rest$ bipartition, as considered in Proposition~1. Moreover, the results considered in \cite{lightcone-prabhu} are limited to the measures, concurrence squared and quantum discord, in contrast to the present results which are valid for all quantum correlation measures.}

{In subsequent sections, we analyze the validity of the statements made in Theorem~\ref{p1} and Proposition~1, through the quantum correlation measures $\cal{C}^2$ and $\cal{N}^2$, from the entanglement-separability paradigm, and $\cal{D}$ and $\Delta$, from the information-theoretic one.
Table~{\ref{table1}} provides the specific 
form of the functions $\cal{F}^{\cal{Q}}(b)$ and $\cal{R}^{\cal{Q}}(b,\beta)$ for the measures, $\cal{C}^2$, $\cal{N}^2$, $\cal{D}$, and $\Delta$.} We refer to the bounds obtained on the monogamy scores as the multiparty entanglement bounds.

\section{Analyzing the bounds for special multiqubit states}
\label{anal}

In this section, we study some important classes of multipartite states for which the multiparty entanglement bound on monogamy score holds.
If in the evaluation of GGM, the maximum eigenvalue is obtained from the $1$:$rest$ bipartition, then monogamy score is always bounded above by the GGM  via Theorem~\ref{p1}. However, Proposition 1 only holds when a state  obeys certain conditions. We consider several paradigmatic states for which we check whether the criteria required for Proposition 1 to hold are satisfied.

\subsection{Dicke States}
\label{dicke}

Let us consider an $n$-qubit Dicke state \cite{dicke} with $r$ excitations, given by the equation
\begin{equation}
|\Psi^n_r\rangle_{D} = \dbinom{n}{r}^{-\frac{1}{2}} \sum \mathcal{P}\left(|0\rangle^{\otimes(n-r)} \otimes |1\rangle^{\otimes r}\right),
\label{dick}
\end{equation}
where the summation is over all possible permutations ($\cal{P}$) of the product state having \(r\) qubits in the excited state, \(|1\rangle\),  and \(n-r\) qubits in the ground state, \(|0\rangle\). The state $|\Psi^n_1\rangle_{D}$ is the well-known W state. Since, the Dicke state is symmetric, all $k$:$rest$ bipartitions are equivalent, and the reduced density matrix can be written as
\begin{equation}
\rho_D^{(k)} = \frac{1}{\binom{n}{r}} \sum_{i=0}^k \binom{k}{i}\binom{n-k}{r-i} |\Psi^k_i\rangle_D\langle\Psi^k_i|.
\label{red-dick}
\end{equation}
Using Eq.~(\ref{red-dick}), 
the maximum eigenvalue contributing to GGM can be obtained. For $r \neq n/2$, the maximum eigenvalue comes from the $1$:$rest$ ($k=1$) bipartition \cite{Bergmann, Toth} and is equal to $a$, where
\begin{eqnarray}
a = \binom{n-1}{r-1}\bigg/\binom{n}{r} &=& \frac{r}{n},~~~~~~~ \mathrm{for}~~ r > \frac{n}{2}, \nonumber\\
\mathrm{and}~~ a = \binom{n-1}{r}\bigg/\binom{n}{r} &=& 1- \frac{r}{n},~ \mathrm{for}~~ r < \frac{n}{2}.
\end{eqnarray}
Hence, for the $n$-qubit Dicke state $|\Psi^n_r\rangle_{D}$ with $r \neq n/2$, the bound on monogamy score is satisfied via Theorem~\ref{p1}.

However, for the Dicke state with  $r = n/2$, the situation is involved. The maximum eigenvalue comes from the $2$:$rest$ ($k$ = $2$) bipartition \cite{Bergmann, Toth} and is equal to $b = 2 \binom{n-2}{r-1}/\binom{n}{r}$ = $\frac{n}{2(n-1)}$. Hence, for $r = n/2$, the quantities $\beta$ = $b-a$ = $\frac{1}{2(n-1)} > 0$,  $\cal{R}^{\cal{C}^2(\cal{N}^2)}$ = $z\beta(1-2b+\beta)$ = $-\frac{z}{4(n-1)^2} < 0$, where $z = 4~ (1)$ for $\cal{C}^2~ (\cal{N}^2)$, and $\cal{R}^{\cal{ D}(\Delta)}=-\frac{1}{2}\left(\log_2{\frac{n(n-2)}{(n-1)^2}} + \frac{1}{n-1}\log_2{\frac{n}{n-2}}\right) < 0$. Hence, for all even \(n \geq 4\) and $r = n/2$, for the monogamy score bound to be satisfied, we must have $\cal{H}^{\cal{Q}}(|\Psi^n_r\rangle_{D}) \geq 0$.

For the symmetric Dicke states, analytical forms for $\cal{C}^2$, $\cal{N}^2$ and $\cal{D}$, for any two-qubit density matrices, are known and can be written as \cite{newhri1}
\begin{eqnarray}
{\cal C}^{2}(\rho^{(2)}_{ij})&=& 4(v-\sqrt{uw})^2, \\
{\cal N}^{2}(\rho^{(2)}_{ij})&=& \frac{1}{4}|(u+w)-\sqrt{(u-w)^2+4v^2}|^2, \\
{\cal D}(\rho^{(2)}_{ij})&=& S'- S'' + h(l), 
\end{eqnarray}
where
$l=\frac12 \left(1+\sqrt{1-4(uv+vw+wu)}\right)$,
$S'=-(u+v)\log_2(u+v)-(v+w)\log_2(v+w)$, 
$S''=-u\log_{2}u-2v\log_{2}2v-w\log_{2}w$, 
and 
$u=(n-r)(n-r-1)/(n^2-n), 
v= r(n-r)/(n^2-n)$, and $w=r(r-1)/(n^2-n)$.

For the Dicke state with  $r = n/2$, all these quantities become functions of a single parameter, the size of the state, $n$. {Using Eq.~(\ref{bound2-p1}) for symmetric pure states}, it can be easily shown that the quantity, $\cal{H}^{\cal{Q}}(|\Psi^n_{n/2}\rangle_{D})$ = $\cal{R}^{\cal{Q}}+(n-1){\cal Q}(\rho^{(2)}_{ij}) \geq 0$, for the quantum correlation measures $\cal{C}^2$, $\cal{N}^2$, and $\cal{D}$. Thus, 
from Proposition 1, the GGM is the upper bound on monogamy scores for these states.

\subsection{Generalized superposition of GHZ and W states}
\label{gengw}

Let us consider the permutationally invariant states defined by a superposition of generalized GHZ state and W state, given by
\begin{equation}
|\Psi^n_{\tilde\alpha,\tilde\gamma}\rangle = \tilde\alpha|0\rangle^{\otimes n} + \tilde\beta|1\rangle^{\otimes n} + \tilde\gamma|W^{n}\rangle,
\end{equation}
where $(\tilde\alpha, \tilde\beta, \tilde\gamma) \in \mathbb{C}$ and $|\tilde\beta| = \sqrt{1-|\tilde\alpha|^2-|\tilde\gamma|^2}$. $|W^{n}\rangle$ is the $n$-qubit W state, and for $\tilde\gamma = 0$, $|\Psi^n_{\tilde\alpha,\tilde\gamma}\rangle$ is the generalized GHZ state. To obtain the reduced density matrices, one can notice that the state can be rewritten in the form
\begin{eqnarray}
|\Psi^n_{\tilde\alpha,\tilde\gamma}\rangle &=& \tilde\alpha |0\rangle^{\otimes k} |0\rangle^{\otimes n-k} + \tilde\beta \sqrt{\frac{n-k}{n}} |0\rangle^{\otimes k} |W^{n-k}\rangle \nonumber\\
&+& \tilde\beta \sqrt{\frac{k}{n}} |W^{k}\rangle |0\rangle^{\otimes n-k} + \tilde\gamma |1\rangle^{\otimes k} |1\rangle^{\otimes n-k}.
\end{eqnarray}
Therefore, the reduced $k$-qubit density matrix can be written as
\begin{equation}
\rho^{(k)}_{\tilde\alpha,\tilde\gamma}= 
\left(\begin{array}{ccc}
|\tilde\alpha|^2+|\tilde\beta|^2\frac{n-k}{n} & \tilde\alpha^*\tilde\beta\sqrt{\frac{k}{n}} & 0 \\
\tilde\alpha\tilde\beta^*\sqrt{\frac{k}{n}} & |\tilde\beta|^2\frac{k}{n} & 0 \\
0 & 0 & |\tilde\gamma|^2 
\end{array}\right),
\label{ghz-w}
\end{equation}
in the orthogonal basis formed by $|0\rangle^{\otimes k}$, $|W^{k}\rangle$, and 
$|1\rangle^{\otimes k}$.
By evaluating the eigenvalues of the above matrix, we find that the maximum eigenvalue corresponds to the $1$:$rest$ ($k=1$) bipartition and is given by
\[
a = \frac{1}{2}\left(1+\sqrt{1-4|\tilde\alpha|^2|\tilde\gamma|^2 + \frac{4(n-1)}{n}|\tilde\beta|^2(|\tilde\gamma|^2 + \frac{|\tilde\beta|^2}{n})}\right).
\]
Hence for $|\Psi^n_{\tilde\alpha,\tilde\gamma}\rangle$, the multiparty entanglement bound on monogamy score is satisfied via Theorem~\ref{p1}.

\begin{figure*}[t]
\begin{center}
\includegraphics[width=16cm, angle=0]{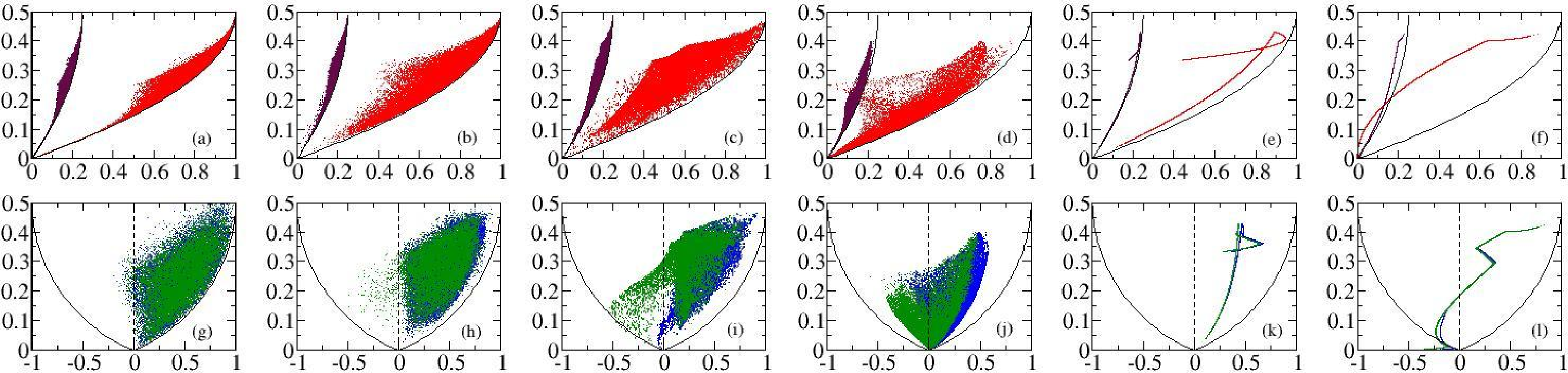}
\caption{(Color online.) Genuine multiparty entanglement versus quantum monogamy scores for the SLOCC inequivalent classes. The figure exhibits plots of quantum monogamy scores ($\delta^{\cal{Q}}$), as the abscissae, against the generalized geometric measure ($\cal{G}$), as the ordinates. Monogamy scores for squared-concurrence (red dots) and squared-negativity (maroon dots) 
are shown in the first row (Figs. 1(a)-1(f)), and quantum discord (blue dots) and quantum work-deficit (green dots) are shown in the second row (Figs. 1(g)-1(h)). Each of the six columns represents plots for $2.5 \times 10^5$ random states, generated through uniform Haar distribution, for the normal-form representatives of the six four-qubit SLOCC inequivalent classes ($|G^1\rangle$ ((a) and (g)) through $|G^6\rangle$ ((f) and (l)) given in Table~\ref{table-9classes}). The multiparty entanglement bounds on the monogamy scores are given by the equation, $\delta^\cal{Q}(|\Psi\rangle) =  \cal{F}^{\cal{Q}}(\cal{G}(|\Psi\rangle))$, as proposed in Sec.~\ref{bound}. Monogamy scores for quantum discord and quantum work-deficit, in the second row, can be negative but are not bounded by the negative of the entropic function, i.e., the mirror image of the equation $\delta^{\cal{Q}} = \cal{F}^{\cal{Q}}(\cal{G})$, for $\delta^{\cal{Q}}>0$, about the $\delta^{\cal{Q}}=0$ axis.
The abscissae are measured in ebits. The ordinates are measured in ebits for Figs. 1(a)-1(f) and in bits for Figs. 1(g)-1(h).
}
\label{figure1}
\end{center}
\end{figure*}

\subsection{The Majumdar-Ghosh model}
\label{mg}
 
Let us now consider a physical system that is useful in studying quantum phenomena in strongly-correlated quantum spin systems. The Majumdar-Ghosh (MG) model \cite{MG-model} is a one-dimensional, antiferromagnetic frustrated system, with a Hamiltonian given by 
\begin{equation}
H_{\mathrm{MG}} = J_1 \sum_{\langle i,j \rangle} \vec{\sigma_i}\cdot\vec{\sigma_j} + \frac{J_1}{2} \sum_{\langle\langle i,j \rangle\rangle} \vec{\sigma_i}\cdot\vec{\sigma_j},
\end{equation}
where $\langle i,j \rangle$ and $\langle\langle i,j \rangle\rangle$ refer to the nearest and the next-nearest neighbors interactions respectively. $\vec{\sigma} = (\sigma^x,\sigma^y,\sigma^z)$ are the Pauli spin operators, and $J_1 >$ 1. Here, we assume that the number of spins, $n$, is even, and the chain is periodic. The MG model is a special case of the more general $J_1-J_2$ model for which the ground state is exactly known for $J_2 = J_1/2$ \cite{MG-model}. 
The $n$-qubit ground state is doubly degenerate and frustrated. The ground state space is spanned by
\begin{equation}
|\psi^n_{(\pm)}\rangle = \frac{1}{2^{n/4}} \prod_{i=1}^{n/2} (|0_{2i}1_{2i\pm1}\rangle - |1_{2i}0_{2i\pm1}\rangle).
\end{equation}
Let us consider the ground state
\begin{equation}
|\Psi^n_{\mathrm{MG}}\rangle = |\psi^n_{(+)}\rangle + |\psi^n_{(-)}\rangle.
\end{equation}
It is known to be genuinely multipartite entangled and is rotationally invariant \cite{MG-model-ground-rotinv}. For $n \geq 4$, the maximum eigenvalue is known to come from the $2$:$rest$ nearest-neighbor bipartition, where the reduced two-qubit density matrix is the rotationally invariant Werner state. The maximum eigenvalue from the $1$:$rest$ bipartition is $a$ = 1/2. The maximum eigenvalue from the nearest-neighbor $2$:$rest$ bipartition is $b = (1+3p)/4$, where $p$ is the Werner parameter, given by 
\begin{equation}
p = \frac{1+2^{\frac{n}{2}-2}}{1+2^{\frac{n}{2}-1}}.
\label{wer}
\end{equation}
Hence for $n > 4$, we have $p > 1/3$, which implies $\beta = b - a = (3p-1)/4 >$ 0. For the reduced two-site density matrix, the exact analytical forms for $\cal{C}^2$, $\cal{N}^2$, and $\cal{D}$ are known in terms of the Werner parameter $p$, and can be written as
\begin{eqnarray}
{\cal C}^{2}(\rho^{(2)}_{ij})&=& \max\left[0,\frac{3p-1}{2}\right]^2, ~~~~
{\cal N}^{2}(\rho^{(2)}_{ij})= \left|\frac{1-3p}{4}\right|^2 ,\nonumber \\
{\cal D}(\rho^{(2)}_{ij})&=& \frac{p_-}{4}\log_2(p_-)-\frac{p_+}{2}\log_2(p_+)+\frac{p'}{4}\log_2(p'), 
\end{eqnarray}
where $p_{\pm} = 1 \pm p$, $p'=1+3p$, and $i$ and $j$ are the nearest-neighbors.
To prove that the multiparty entanglement bounds on monogamy scores hold for these quantum correlations for the ground states of the MG model, we need to show that the quantity, $\cal{H}^\cal{Q}(|\Psi^n_{\mathrm{MG}}\rangle)$, is positive. Using Eq.~(\ref{wer}), we can derive that $\cal{R}^{\cal{C}^2(\cal{N}^2)}$ = $-z\left(\frac{1-3p}{4}\right)^2$, for $p > 1/3$. Similarly, one can derive the expression for $\cal{R}^{\cal{D}(\Delta)}$. We need to prove that the quantity $\cal{H}^\cal{Q}(|\Psi^n_{\mathrm{MG}}\rangle) >$ 0. For the ground state, $|\Psi^n_{\mathrm{MG}}\rangle$, only the nearest-neighbor spins are entangled and ${\cal C}^{2}(\rho^{(2)}_{ij})= {\cal N}^{2}(\rho^{(2)}_{ij}) = 0$, for $j \neq i\pm1$. ${\cal D}(\rho^{(2)}_{ij})$ for non-nearest-neighbor qubits is finite but two orders of magnitude lower than the nearest-neighbor values. Hence, {using Eq.~(\ref{bound2-p1}), one obtains} $\cal{H}^\cal{Q}(|\Psi^n_{\mathrm{MG}}\rangle) = \cal{R}^\cal{Q} + {\cal Q}(\rho^{(2)}_{i(i+1)})+{\cal Q}(\rho^{(2)}_{i(i-1)})$,  {where we approximate ${\cal D}(\rho^{(2)}_{ij})= {\Delta}(\rho^{(2)}_{ij}) \approx 0$, for $j \neq i\pm1$}. For $\cal{C}^2$ and $\cal{N}^2$, $\cal{H}^{\cal{C}^2(\cal{N}^2)}$ = $\frac{z}{16} (1-3p)^2 >$ 0. Similarly, for $\cal{D}$ and $\Delta$, one can show that $\cal{H}^{\cal{D}(\Delta)} >$ 0, for all $n$. Hence, the monogamy score bound is satisfied via Proposition 1.

\subsection{The Ising model}
\label{ising}

In this section, we consider two paradigmatic Hamiltonians belonging to the Ising group of models \cite{bikas} that give us multipartite entangled ground states. We first consider the highly frustrated Ising model with long-range antiferromagnetic interactions, also called the Ising gas model. The Hamiltonian for an $n$-spin Ising gas is given by
\begin{equation}
H_{\mathrm{gas}}(x) = \frac{J}{n}(S - nx)^2,~~~ J > 0,
\end{equation}
where $S$ = $\sum_i \sigma_i^z$, $J >$ 0, and 0 $\leq x \leq$ 1. The quenched unnormalized ground state of the Ising gas Hamiltonian is given by \cite{MG-model}
\begin{equation}
|\Psi^n_{\mathrm{gas}}\rangle = \sum_{\left\{[0,1]\right\}} |0\rangle^{\otimes n(1+x)/2} \otimes |1\rangle^{\otimes n(1-x)/2},
\end{equation}
where $\left\{[0,1]\right\}$ indicates that the summation is over all possible combinations of $|0\rangle$ and $|1\rangle$ that satisfy the density $(1+x)/(1-x)$. For maximally frustrated ground states, the density is unity ($x = 0$), and the ground state reduces to the Dicke state, given by Eq.~(\ref{dick}), for $r = n/2$. For these states, as discussed in Sec.~\ref{dicke}, the multiparty entanglement always gives the upper bound on the monogamy of quantum correlation. 

We next consider the weakly frustrated, periodic Ising spin chain with nearest-neighbor interactions, also called the Ising ring. All interactions are ferromagnetic, except one that is antiferromagnetic. The Hamiltonian is given by
\begin{equation}
H_{\mathrm{ring}} = -J\sum_{i=1}^{n-1} \sigma^z_i \sigma^z_{i+1} + J \sigma^z_n \sigma^z_{1},~~~ J > 0.
\end{equation} 
The quenched ground state of the Hamiltonian is given by \cite{MG-model}
\begin{eqnarray}
|\Psi^n_{\mathrm{ring}}\rangle &=& \sum_{k=0}^{n-1} \Big(|0^{\otimes n-k} 1^{\otimes k}\rangle + |1^{\otimes n-k} 0^{\otimes k}\rangle  
+ \left. |1^{\otimes k+1} 0^{\otimes n-k-1)} \rangle + |0^{\otimes k+1} 1^{\otimes n-k-1)} \rangle \right).\,\,\,\,\,
\label{ring}
\end{eqnarray}
For the ground state given in Eq.~(\ref{ring}), the reduced density matrix from the $2$:$rest$ bipartition can be written as
\begin{equation}
\rho^{(2)}_{\mathrm{ring}}= \frac{1}{2n}
\left(\begin{array}{cccc}
n-1 & 1 & 1 & 2 \\
1 & 1 & 0 & 1 \\
1 & 0 & 1 & 1 \\
2 & 1 & 1 & n-1 
\end{array}\right).
\label{densering}
\end{equation}
The maximum eigenvalue for the $2$:$rest$ bipartition is given by $b$ = $\frac{1}{4n}(n+2+\sqrt{n^2+16})$, and for the $1$:$rest$ bipartition is given by $a$ = $(1/2)(1+1/n)$. These eigenvalues are highest among all bipartitions.
For any finite number of spins $n$, $a \ge b$. 
Therefore, the bound on monogamy is satisfied via Theorem~\ref{p1}.
Interestingly, for $n \rightarrow \infty$, the maximum eigenvalues, $a = b = 1/2$, and maximum GGM is achieved.



%
\begin{figure}
\begin{center}
\includegraphics[width=7cm, angle=0]{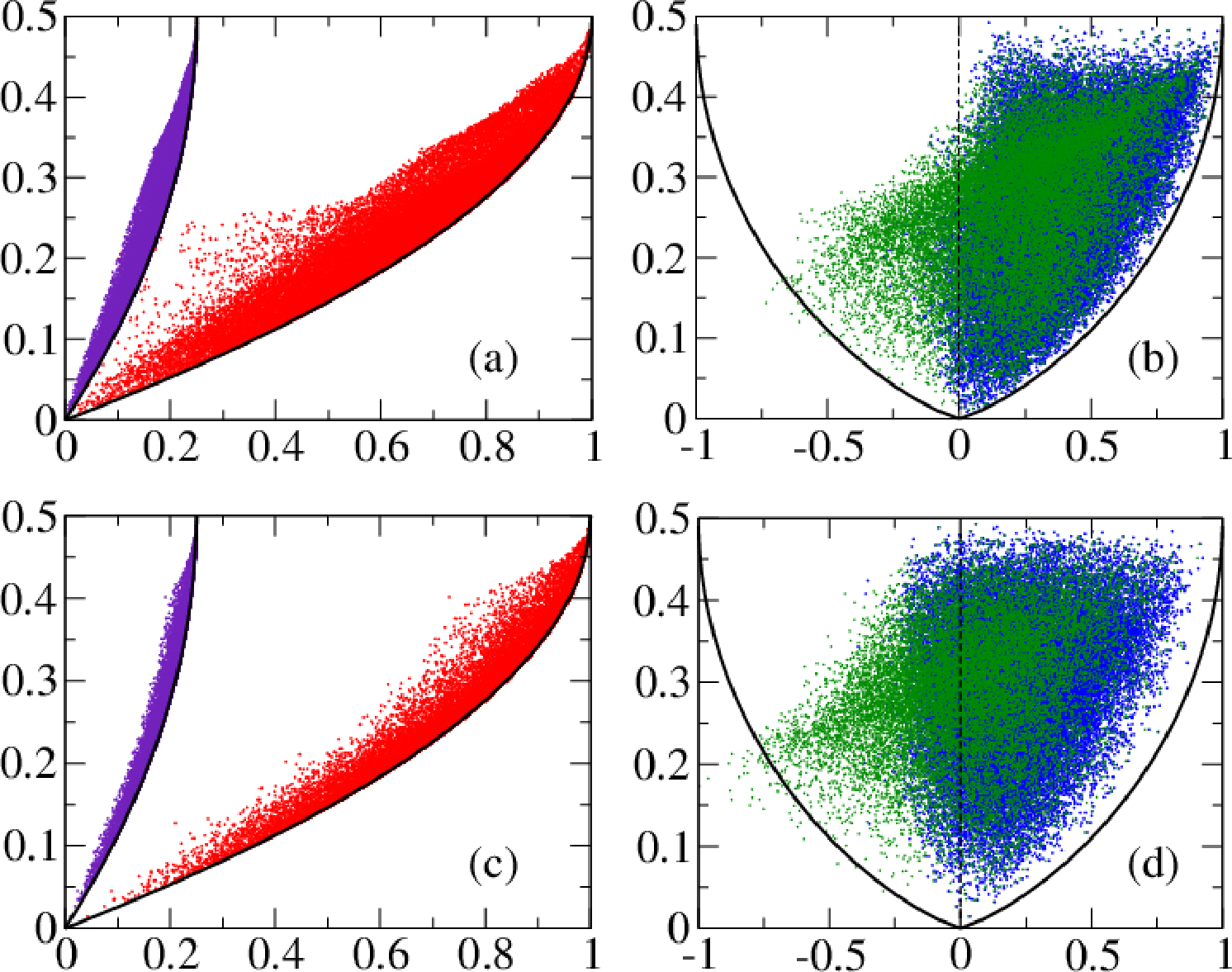}
\caption{(Color online.) Genuine multiparty entanglement versus quantum monogamy scores for symmetric states. The figure shows the plot of quantum monogamy scores, along the abscissae, against the generalized geometric measure, along the ordinates, for symmetric four-qubit (Figs.~2(a) and 2(b)) and five-qubit (Figs. 2(c) and 2(d)) states generated using random superposition of Dicke states, as shown in \ref{symmetric}. The description of quantum correlation measures, random state generation, bounds, and the axes  are same as in Fig.~\ref{figure1}.
}
\label{figure2}
\end{center}
\end{figure}
%

\section{Numerical Results}
\label{num}

The $n$-qubit states, considered in the analytical study of the bound on quantum monogamy in the previous section, constitute 
some special classes of multiparty state of arbitrary number of qubits.
To visualize the multiparty entanglement bound obtained in Theorem~\ref{p1} and Proposition~1, we now
randomly generate four- and five-qubit states. The random states are chosen using Haar uniform distribution.

Fig.~\ref{figure1} depicts the behavior of genuine multiparty entanglement with respect to the quantum monogamy scores
for $\cal{C}^2$ and $\cal{N}^2$ (Figs. 1(a)-1(f)) and for $\cal{D}$ and $\Delta$ (Figs. 1(g)-1(l)) for the randomly generated four-qubit states corresponding to the parametrized six SLOCC inequivalent classes ($|G^i\rangle$, for $i$ = 1 to 6). The nine SLOCC inequivalent classes of four-qubit states are discussed in \ref{four} and their exact forms are given in Table~\ref{table-9classes}.  
%
%

Fig.~\ref{figure1} shows that the quantum monogamy scores are bounded by the quadratic and entropic functions of generalized geometric measure for the set of states belonging to the SLOCC inequivalent classes for four-qubits. It is known that quantum discord and quantum work-deficit can have negative monogamy scores for certain states, i.e., the measures are not monogamous \cite{viola,viola1, viola3}. This is observed by the negative regions Figs.~\ref{figure1}(g)- \ref{figure1}(l) .

\begin{figure}
\begin{center}
\includegraphics[width=7cm, angle=0]{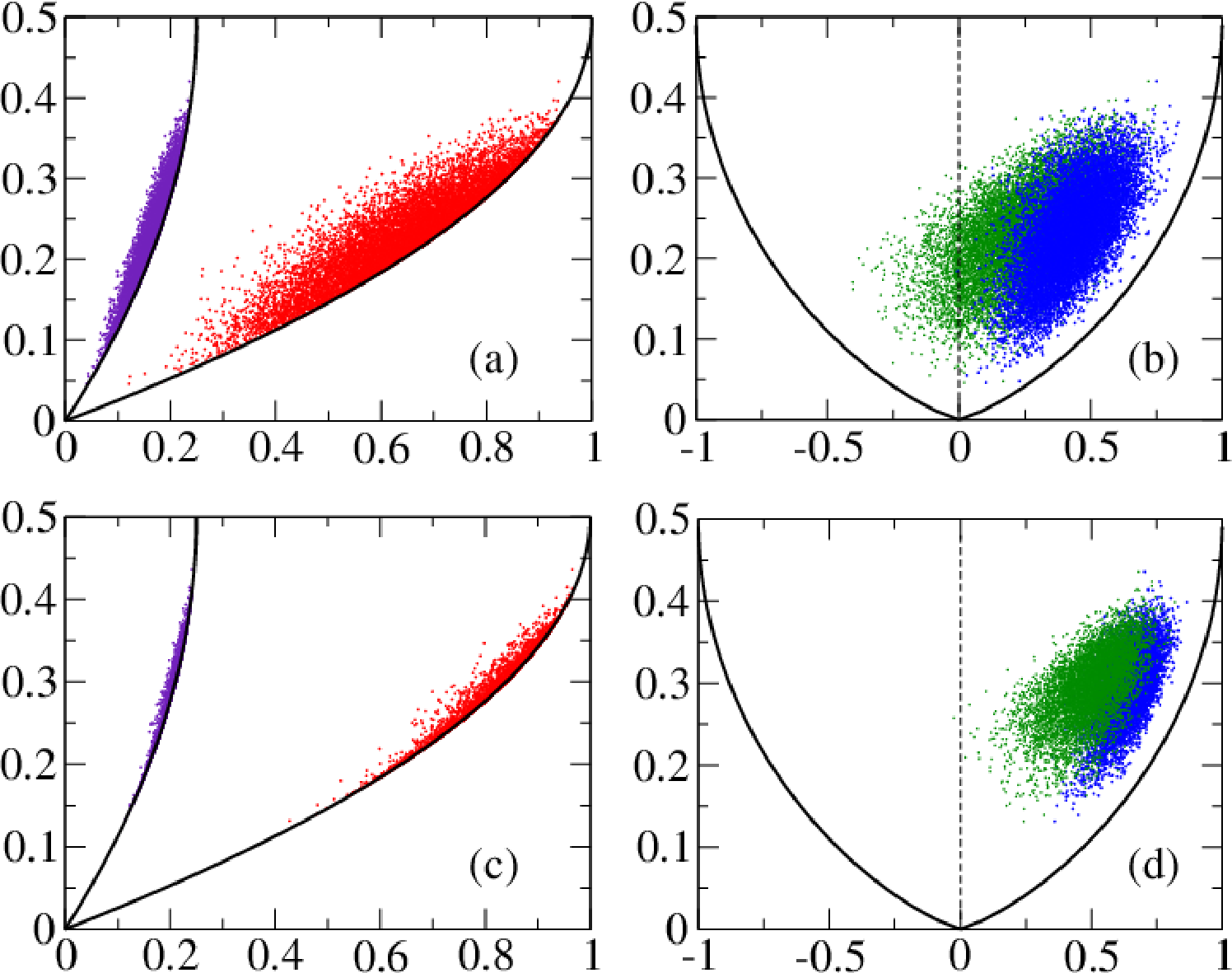}
\caption{(Color online.) Genuine multiparty entanglement versus quantum monogamy scores for generic four- and five-qubit states. The description is same as in Fig.~\ref{figure2}.
}
\label{figure3}
\end{center}
\end{figure}

Fig.~\ref{figure2}  shows the bound on monogamy scores for randomly generated symmetric four- and five-qubit states. The symmetric states are generated using a random superposition of Dicke states, with different excitations, as shown in \ref{symmetric} 
by using a uniform Haar distribution. The figure shows that the generated symmetric four- and five-qubit states satisfy the multiparty entanglement bound on quantum monogamy in terms of the functions of the generalized geometric measure. Fig.~\ref{figure3} exhibits the bound for randomly generated four- and five-qubit states 
using a uniform Haar distribution.

From the analytical and numerical results obtained in the preceding and this sections,
it is observed that the bound on quantum monogamy scores for the quantum correlation measures $\cal{C}^2$, $\cal{N}^2$, $\cal{D}$, and $\Delta$, in terms of derived functions of the generalized geometric measure, is satisfied for a large majority of multiqubit quantum states.

\section{Conclusion}
\label{conclude}

Monogamy is an intrinsic feature of quantum correlation that distinguishes it from classical correlations and plays an important role in applications of quantum information theory such as quantum cryptography and other multiparty communication protocols. 
While monogamy provides a significant advantage in obtaining bounds on secret key rates in quantum cryptography, it also restricts the availability of resources in multipartite protocols such as quantum dense coding. Thus, the monogamy score is an important aspect in the study of various quantum information protocols, as it shows the restrictions posed by quantum mechanics in a multiparty domain.
However, the monogamy score is a difficult quantity to compute and estimate for generic quantum states and for arbitrary measures of quantum correlations. It is therefore interesting to derive bounds on the monogamy score. This would allow a better understanding of the constraints on quantum correlations in many-body systems even when the exact monogamy inequalities are not accessible. Our work provides an easily estimable upper bound on the monogamy score.


In this work, we find that the monogamy score of any quantum correlation measure, for a large majority of $n$-qubit quantum states, is bounded above by certain simple functions of the generalized geometric measure, which quantifies the amount of genuine multipartite entanglement present in the system. {We find that the bound is 
universally satisfied for all three-qubit states.} 
We show that such an upper bound holds also for an arbitrary number of qubits provided the states satisfy certain conditions. 
We derive a set of necessary conditions to characterize the set of states that may violate the bound, and numerically observe that the set is extremely small. Moreover, we analytically investigate several important classes of multiparty quantum states with arbitrary number of parties for which we show that the conditions required to have the upper bound on monogamy scores of computable bipartite measures are satisfied.

The obtained monogamy score bound due to the genuine multiparty entanglement in the system shows a forbidden regime in the distribution of bipartite quantum correlation measures  among different parties in a multiparty system.
{More specifically, given a certain amount of genuine multipartite entanglement, the monogamy score corresponding to any bipartite quantum correlation measure is forbidden to go above a certain value.}
The results provide a unifying framework to study monogamy relations in both entanglement and information-theoretic quantum correlations.

\section*{Acknowledgements}
HSD acknowledges funding by the Austrian Science Fund (FWF), project no.~M 2022-N27, under the Lise Meitner programme
of the FWF.
RP acknowledges the INSPIRE-faculty position at Harish-Chandra Research Institute (HRI) from the Department of Science and Technology, Government of India. The authors acknowledge computations performed at the cluster computing facility at HRI.

\appendix
\section*{APPENDIX}

\begin{table*}[t]
\begin{center}
\begin{tabular}{|l|}
\hline
\vspace{0.01cm}
$~|G^x\rangle$ (unnormalized) \\ 
\hline
$~|G^1_{abcd}\rangle=\frac{a+d}{2}(|0000\rangle + |1111\rangle) + \frac{a-d}{2}(|0011\rangle + |1100\rangle) + \frac{b+c}{2}(|0101\rangle + |1010\rangle) + \frac{b-c}{2}(|0110\rangle + |1001\rangle) $ \\~\\
$~|G^2_{abc}\rangle=\frac{a+b}{2}(|0000\rangle + |1111\rangle) + \frac{a-b}{2}(|0011\rangle + |1100\rangle) + c(|0101\rangle + |1010\rangle) + |0110\rangle$\\~\\
$~|G^3_{ab}\rangle=a(|0000\rangle + |1111\rangle) + b(|0101\rangle + |1010\rangle) + |0110\rangle + |0011\rangle$\\~\\
$~|G^4_{ab}\rangle=a(|0000\rangle + |1111\rangle) + \frac{a+b}{2}(|0101\rangle + |1010\rangle) + \frac{a-b}{2}(|0110\rangle + |1001\rangle)$\\~\\
$~~~~~~~~~~~+ \frac{i}{\sqrt{2}}(-|0001\rangle - |0010\rangle + |0111\rangle + |1011\rangle)$\\~\\
$~|G^5_{a}\rangle=a(|0000\rangle +|0101\rangle + |1010\rangle + |1111\rangle) + i(|0001\rangle - |1011\rangle)+ |0110\rangle$\\~\\
$~|G^6_{a}\rangle=a(|0000\rangle + |1111\rangle) + |0011\rangle + |0101\rangle + |0110\rangle$\\~\\
$~|G^7\rangle=|0000\rangle + |0101\rangle) + |1000\rangle + |1110\rangle$\\~\\
$~|G^8\rangle=|0000\rangle + |1011\rangle) + |1101\rangle + |1110\rangle$\\~\\
$~|G^9\rangle=|0000\rangle + |1111\rangle$  \\\hline
\end{tabular}
\caption{Normal-form representatives of the nine four-qubit SLOCC inequivalent classes defined in \cite{verstraete-9classes,chterental} (see \ref{four}). Here \(a\), \(b\), \(c\), \(d\) are complex parameters with non-negative real parts. The first six classes are parameterized.}
\label{table-9classes}
\end{center}
\end{table*}

\subsection{Four-qubit SLOCC classes}
\label{four}
For three-qubit pure states, there exist two inequivalent classes of states under stochastic local operation and classical communication (SLOCC), namely the $GHZ$ and the $W$ class of states \cite{threeslocc}. However, for four-qubits, there exist infinitely many inequivalent SLOCC classes of states \cite{siewert}. 
%
A useful classification into nine classes for four-qubit was obtained in \cite{verstraete-9classes,chterental}. It was observed that up to permutation of the qubits, any four-qubit pure state can be transformed into one of the nine classes of states $\{|G^x\rangle \}$, as shown in Table~\ref{table-9classes}. 

\subsection{Symmetric States}
\label{symmetric}
A general symmetric state can be written as a linear combination of the Dicke states as 
\begin{equation}
|\Psi_S\rangle=\sum_{r=0}^n a_r~|\Psi_r^n\rangle_D,
\end{equation}
where $|\Psi_r^n\rangle_D$ is an $n$-qubit Dicke state \cite{dicke} with $r$ excitations, given in Eq.~(\ref{dick}). The normalization condition is satisfied by demanding $\sum_{r=0}^n|a_r|^2=1$. 
Any general symmetric state can be generated by randomly choosing a set of coefficients $a_r$ that satisfy the normalization. 
\subsection{Quantum correlation measures}

\noindent\emph{Concurrence}: Concurrence \cite{concurrence} is a useful measure of entanglement for general two-qubit states. The concurrence of any two-qubit state, \(\rho_{AB}\), is given by, 
\begin{equation}
{\cal C}(\rho_{AB})=\mbox{max}\{0,\lambda_1-\lambda_2-\lambda_3-\lambda_4\},
\end{equation}
where $\lambda_i$'s are the square roots of the eigenvalues of $\rho_{AB}\tilde{\rho}_{AB}$, arranged in a decreasing order.
$\sigma_y$ is the Pauli spin matrix and $\tilde{\rho}_{AB}=(\sigma_y\otimes\sigma_y)\rho_{AB}^*(\sigma_y\otimes\sigma_y)$, with the complex conjugation being done in the computational  basis.
The concurrence of a two-qubit pure state, $|\Psi\rangle_{AB}$, reduces to $2\sqrt{\det \rho_A}$, where ${\rho_A} = \textrm{Tr}_B(|\Psi\rangle\langle\Psi|_{AB})$. 
%
The concurrence for arbitrary two-qubit states can be used to derive a closed form of the entanglement of formation \cite{eof}, as shown in \cite{concurrence}.


\noindent\emph{Negativity}:  Another important and computable measure of entanglement is negativity \cite{negativity}. \({\cal N}(\rho_{AB})\), of a bipartite state \(\rho_{AB}\) is defined as the sum of the absolute values of the negative eigenvalues of \(\rho_{AB}^{T_{A}}\), which is the partial transpose of \(\rho_{AB}\) with respect to subsystem \(A\). 
Mathematically, \({\cal N}(\rho_{AB})\) can be expressed as
\begin{equation}
{\cal N}(\rho_{AB})=\frac{\|\rho_{AB}^{T_A}\|_1-1}{2},
\end{equation}
where $\|\rho_{AB}^{T_A}\|_1$ 
is the trace-norm of the matrix $\rho_{AB}^{T_A}$.
For two-qubit states, zero negativity implies that the state is separable. 


\noindent\emph{Quantum discord}: In classical information theory, the mutual information between two random variables is given by the following two equivalent expressions:
\begin{eqnarray}
\mathcal{I}(A:B) = H(A)+ H(B)- H(A,B),~~~ 
{\cal J}(A:B) = H(B) - H(B|A), 
\label{cmi}
\end{eqnarray}
where $H(\cdot)$ is the Shannon entropy \cite{ShannonEntropy}. 
For quantum systems, using von Neumann entropy \cite{vonNeumannEntropy} instead of Shannon entropy, one obtains, for a bipartite state $\rho_{AB}$, the expressions
\begin{eqnarray}
\mathcal{I}(\rho_{AB}) = S(\rho_A)+ S(\rho_B)- S(\rho_{AB}),~~~
{\cal J}(\rho_{AB}) = S(\rho_B) - S(\rho_{B|A}), 
\label{qmi}
\end{eqnarray}
where the conditional entropy, 
$
S(\rho_{B|A}) = \min_{\{A_i\}} \sum_i p_i S(\rho_{B|i}),
$
for the state \(\rho_{AB}\),   
with \(\rho_{B|i} = \frac{1}{p_i} \mbox{tr}_A[(A_i \otimes \mathbb{I}_B) \rho (A_i \otimes \mathbb{I}_B)]\),
\(p_i = \mbox{tr}_{AB}[(A_i \otimes \mathbb{I}_B) \rho (A_i \otimes \mathbb{I}_B)]\),
\(\mathbb{I}\) being the identity operator on the Hilbert space of \(B\), and $\{A_i\}$ forms a rank-one projection measurement on the system held by $A$.
The difference between $\mathcal{I}(\rho_{AB})$ and 
${\cal J}(\rho_{AB})$, for a bipartite state 
\(\rho_{AB}\), gives us a measure of quantum correlation of $\rho_{AB}$. 
Quantum discord is defined as \cite{discord1,discord2}
\begin{equation}
\label{eq:discord}
{\cal D}(\rho_{AB})= {\cal I}(\rho_{AB}) - {\cal J}(\rho_{AB}).
\end{equation} 


\noindent\emph{Quantum work-deficit}: Another information-theoretic measure of quantum correlation is quantum work-deficit \cite{qwd1}, which is defined, for a bipartite quantum state \(\rho_{AB}\), as the difference between the quantity of pure states that can be extracted under allowed ``closed global operations'' (CGO) and pure product states that can be extracted under ``closed local operations and classical operations'' (CLOCC).

For a given state $\rho_{AB}$, the class of CGO are any allowed sequences of unitary operations and dephasing using a 
set of projectors $\{\Pi_i\}$, i.e., $\rho \rightarrow \sum_i \Pi_i \rho_{AB} \Pi_i$,  
where $\Pi_i\Pi_j = \delta_{ij} \Pi_i$, $\sum_i \Pi_i = \mathbb{I}$. 
The number of pure qubits that can be extracted from $\rho_{AB}$ by CGO is 
\[I_G (\rho_{AB})= N - S(\rho_{AB}),\]
where $N = \log_2 (\dim {\cal H})$.
The CLOCC class consists of local unitary, 
local dephasing, and exchange of dephased states between $A$ and $B$. The amount of qubits that can be extracted under CLOCC is given by
\begin{equation}
I_L(\rho_{AB}) = N - \inf_{\Lambda \in CLOCC} [S(\rho{'}_{AB})],
\end{equation}
where $\rho{'}_{AB} =\sum_{i} p_i  (A_i\otimes \mathbb{I}_B) \rho_{AB} (A_i\otimes \mathbb{I}_B)$ if one restricts to one-way CLOCC. 
Quantum work-deficit is then defined as
\begin{equation}
\Delta(\rho_{AB}) = I_G(\rho_{AB}) - I_L(\rho_{AB}).
\end{equation}
For such instances, the work-deficit is equal to quantum discord for bipartite states with maximally mixed marginals.

\end{document}